\documentclass[conference]{IEEEtran}
\IEEEoverridecommandlockouts

\newcommand\finline[3][]{\begin{myfont}[#1]{#2}#3\end{myfont}}%
\newenvironment{myfont}[2][]{\csname#2\endcsname[#1]}{}

\usepackage{xspace} 
\newcommand{\middleware}{\textsc{\finline[\fontsize{8}{10}]{myFont}{ADROITNESS}}\xspace}

\newcommand{\enfasis}[1]{\emph{\textbf{#1}}\xspace}
\usepackage{hyperref}

\usepackage{color}

\usepackage{listings}
\definecolor{keywordcolor}{rgb}{0.13,0.13,1}
\definecolor{greenComments}{RGB}{63,127,95}
\lstdefinelanguage{ADL}{
    keywordstyle=[2]\color{purple},
    keywordstyle=[3]\color{blue},
    keywords=[2]{boolean, long, Thread, String, double, equals, false, true, to},
    keywords=[3]{System, Component, Connector, Port, Property, Properties, Role, Roles, Ports, Attachments, Attachment, Heuristic, IF, THEN, ELSE, ELSE-IF, RULE, AND},
    morekeywords={to}
    sensitive=false,
    morestring=[b]",
    morecomment=[l]{//},
    }

\lstdefinestyle{adl-style}{
    language=ADL,
    backgroundcolor=\color{white},   
    commentstyle=\color{green},
    numberstyle=\tiny\color{gray},
    basicstyle=\ttfamily\scriptsize,
    breakatwhitespace=false,         
    breaklines=true,                 
    captionpos=b,                    
    keepspaces=true,                 
    showspaces=false,                
    showstringspaces=false,
    showtabs=false,                  
    tabsize=2,
    columns=fullflexible,
    commentstyle=\color{greenComments}, 
}

\lstdefinestyle{rule-style}{
    language=ADL,
    backgroundcolor=\color{white},   
    commentstyle=\color{greenComments},
    keywordstyle=\color{blue},
    numberstyle=\tiny\color{gray},
    basicstyle=\ttfamily\scriptsize,
    breakatwhitespace=false,         
    breaklines=true,                 
    captionpos=b,                    
    keepspaces=true,                 
    numbers=left,                    
    numbersep=5pt,                  
    showspaces=false,                
    showstringspaces=false,
    showtabs=false,                  
    tabsize=2,
    columns=fullflexible,
    rulecolor=\color{black},
    frame=shadowbox,
    numbers=none
}

\lstnewenvironment{adl-listing}[1][]{
    \lstset{style=adl-style, caption=#1}
  }
  {}
  
\lstnewenvironment{rule-listing}[1][]{
    \lstset{style=rule-style, caption=#1}
  }
  {}  

\usepackage[table]{xcolor}
\definecolor{Gray}{gray}{0.9}

\usepackage{cite}
\usepackage{xspace} 
\usepackage{amsmath,amssymb,amsfonts}
\usepackage{algorithmic}
\usepackage{graphicx}
\usepackage{textcomp}
\usepackage{xcolor}
\def\BibTeX{{\rm B\kern-.05em{\sc i\kern-.025em b}\kern-.08em
    T\kern-.1667em\lower.7ex\hbox{E}\kern-.125emX}}
\begin{document}


\title{Adroitness: An Android-based Middleware for Fast Development of High-performance Apps}

\author{\IEEEauthorblockN{Oscar J. Romero}
\IEEEauthorblockA{\textit{Machine Learning Department} \\
\textit{Carnegie Mellon University}\\
Pittsburgh, USA\\
oscarr@andrew.cmu.edu}
\and
\IEEEauthorblockN{Sushma A. Akoju}
\IEEEauthorblockA{\textit{Machine Learning Department} \\
\textit{Carnegie Mellon University}\\
Pittsburgh, USA\\
sakoju@andrew.cmu.edu}
}

\maketitle

\begin{abstract}
As smartphones become increasingly more powerful, a new generation of highly interactive user-centric mobile apps emerge to make user's life simpler and more productive. Mobile phones applications have to sustain limited resource availability on mobile devices such as battery life, network connectivity while also providing better responsiveness, lightweight interactions within the application. Developers end up spending a considerable amount of time dealing with the architecture constraints imposed by the wide variety of platforms, tools, and devices offered by the mobile ecosystem, thereby diverting them from their main goal of building such apps. Therefore, we propose a mobile-based middleware architecture that alleviates the burdensome task of dealing with low-level architectural decisions and fine-grained implementation details. We achieve such a goal by focusing on the separation of concerns and abstracting away the complexity of orchestrating device sensors and effectors, decision-making processes, and connection to remote services, while providing scaffolding for the development of higher-level functional features of interactive high-performance mobile apps. We demonstrate the powerfulness of our approach vs. Android's conventional framework by comparing different software metrics.

\end{abstract}


\section{Introduction}

Mobile devices, services, and applications have been broadly adopted in everyday activities by users ranging from the business to entertainment domains with increasing demand. Mobile devices come with many a challenge such as limited resources (battery life, network connectivity etc), extending challenge to provide better responsiveness has become the quintessential goal for application architects, developers. \cite{yang2013testing} This ongoing evolution of mobile computing has led developers to develop larger, complex applications that increase the need for methods of reducing software complexity for developing large-scale mobile apps. ~\cite{dehlinger:2011}. The Android Platform, the most used mobile platform by developers and users~\cite{Gartner:2017}, provides a software stack comprising of several app building blocks that allow development of production-quality apps. Developing an Android app involves frequent implementation of new app features, app feature enhancements, fixing domain-specific bugs, refactoring, code clean up activities which constitute 48\% of total development activities of a typical Android developer. \cite{Pascarella:2018:SAA:3197231.3197251} Additionally, consistently keeping up with frequent Mobile platform-level changes and also developing apps for other Mobile platforms only increases the development challenges.\cite{joorabchi2013real} However, acquiring a deep and proper understanding of the Android SDK requires a considerable amount of time for developers (approximately 2+ years~\cite{VisionMobile:2010}) due to the inherent complexity imposed by the over-engineered Android Java Framework (AJF), thereby deviating developers from their main goal: developing robust, interactive, high-performance, production-quality mobile apps. A common way to deal with AJF's complexity is the use of application framework middlewares. These distinguish from Android's native middleware which sits between the specialized Dalvik VM and the operating system layer and includes libraries for many functions including but not limited to data storage, graphics rendering, and web browsing that are compiled to machine language. Such an application framework middleware abstracts the underlying complexity of the development environment and masks the heterogeneity of networking technologies to facilitate app programming~\cite{sanaei:2014}.  Although there exist several such middlewares for Android that range from incorporating context awareness, application awareness, and user awareness including user preferences and behavioral history, they have a significant performance footprint, imply that developers learn additional and complex architectural models yet fail to reduce app development complexity. 

\begin{figure*}[t]
\includegraphics[width=\textwidth]{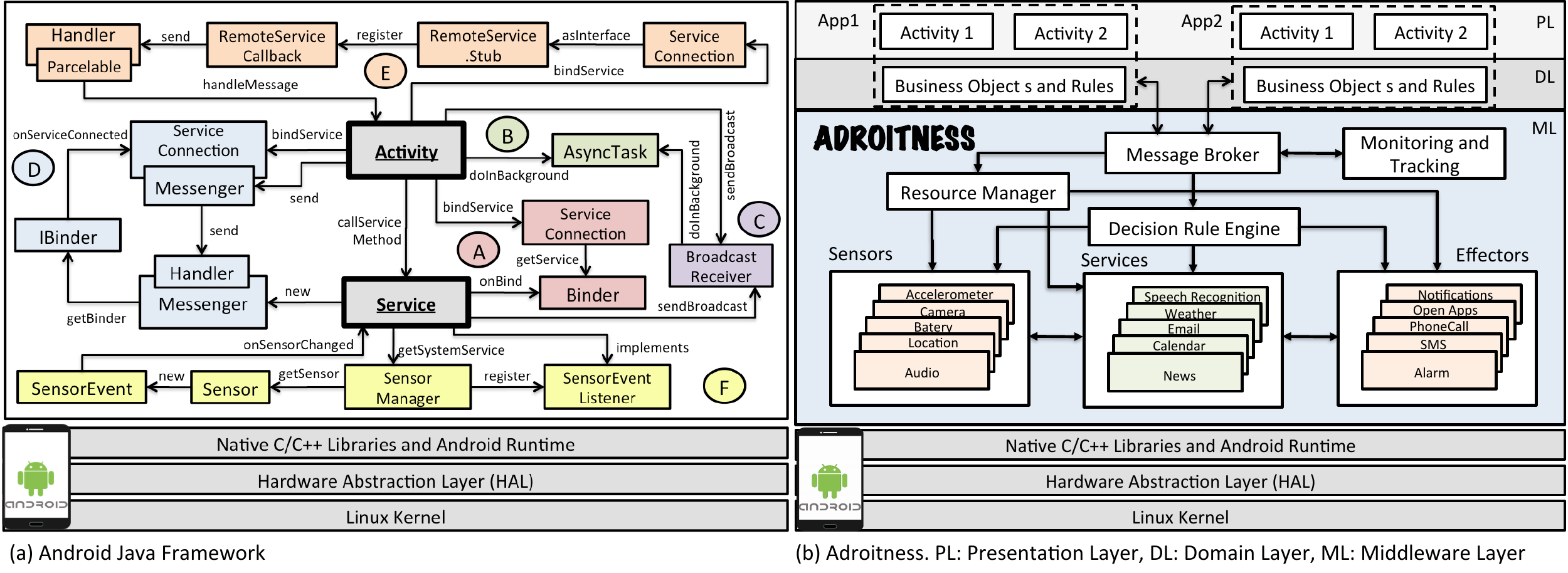}
\vspace{-0.7cm}
\caption{Android Java Framework (AJF) vs. Adroitness Architectural Model}
\vspace{-0.3cm}
\label{fig_architecture}
\end{figure*}

In this paper, we present \middleware, an Android-based middleware architecture that (i) simplifies the implementation of interactive apps by providing programming abstractions from low-level Android app building blocks, (ii) provide a paradigm to avoid writing hundreds of lines of boilerplate code for each new application, and (iii) improve performance of already developed apps by minimal refactoring while meeting certain architectural significant requirements. In section~\ref{sec_motivation} we present our motivation for this work related to existing work, in section~\ref{sec_architecture} we present our architectural model that underlies the development of the proposed Android-based middleware; in section~\ref{sec_implementation} we describe the implementation details of the middleware; in section~\ref{sec_evaluation} we present a comparison between our approach and the Android app building blocks as in AJF which is currently used by most developers, and in section~\ref{sec_conclusions} we summarize our conclusions and present future directions on our research.

\section{Motivation}
\label{sec_motivation}

\subsection{Issues with Android Java Framework (AJF)}
One of the main issues with Android is that it imposes some design and implementation constructs for components to interact with each other, so the resulting apps becomes top-heavy and over-engineered. Another issue with Android is its concurrency architecture, where invoking a simple network request can be a minefield of subtle problems for which even developers with substantial mobile and Java experience may not be prepared~\cite{Mednieks:2013}. 
To illustrate these issues, lets consider an app that sends a network request to a remote service. This simple action spawns several architectural considerations that have to be addressed: 
1) Android modifies the user interface and handles input events from one single thread (the main thread) so any task that occupies it for any significant period of time will cause the UI (Android \enfasis{Activity}) to become unresponsive; 
2) the background tasks should be performed using any of the following Android components: a \enfasis{Handler} (it provides a channel to send data to the main thread), an \enfasis{AsyncTask} (it manages short background asynchronous operations that run on a different thread, so developers would have to deal with thread-safe references and synchronization), a \enfasis{Service} (it has to spawn its own thread in which to do long-running work), or a Java Thread (in this case developers are completely responsible for managing the concurrency). The decision of which kind of component to use depends on several criteria, imposing strong constraints that cannot be verified automatically, e.g., is the background process tied to the UI? is this a long-lasting process? can the process be affected by the activity's lifecycle? is the process shared by multiple components?
3) Android Services define cumbersome mechanisms to communicate with each other, e.g., it is necessary to implement \enfasis{Handlers}, \enfasis{ServiceConnections} (an interface for monitoring the state of a Service), \enfasis{Messengers} (implementation of message-based communication across processes), \enfasis{Intents} (an abstract description of an operation to be performed), \enfasis{IPC} (inter-process communication) etc.;
and 4) Services and Activities can share data across process boundaries by passing \enfasis{Bundles}, objects that implement Serializable or Parcelable interfaces. This process of continuous serialization/deserialization has a significant performance footprint and requires developers to manually parse all the content of these bundles.

\subsection{Requirements:} 
Our work mainly focuses on the following requirements: 
\begin{enumerate}
    \item the middleware should significantly decrease the amount of effort (person/day) and functional size in comparison to an app developed using the AJF.
    \item it must be latency-sensitive, more specifically, responses should not take longer than 100ms (events that complete in 100ms or less are believed to have imperceptible latency and do not contribute to user dissatisfaction~\cite{Shneiderman:2010}).
    \item it must abstract away the complexity of underlying layers (e.g., communication, concurrency, etc.).
    \item it must provide mechanisms for developers to make their apps more modular, pluggable, and easily extensible.
    \item it must provide any kind of mechanism for reasoning over the data collected by the smartphone's sensors and services.
\end{enumerate}
%
%
\section{Adroitness Architectural Model} \label{sec_architecture}
Figure~\ref{fig_architecture} illustrates an architectural model comparison between AJF vs. \middleware in the development of a conventional mobile app. In Figure~\ref{fig_architecture}.a, we have identified 6 different scenarios for Activities and Services to communicate with each other: 
A) the Service is merely a local background worker running in the same process as the Activity, so developer has to create a Binder class and return it to the Activity so it 
In \middleware, developer need to simply have to extend from \middleware 's GenericService, which abstracts away the Android Service implementation, and hence enables developer to focus business logic, feature implementations. \ref{sec_serviceOriented}
B) the Activity uses an AsyncTask to perform background operations and publish results on the UI thread without having to directly manipulate threads and/or handlers; 
\middleware uses service-oriented architecture combined with Message Broker and Resource Locator via Channel Adapters to interact between Activities and Services.
C) Both Activities and Services communicate to each other by sending and receiving broadcast messages through a BroadcastReceiver; \middleware ,  as we describe in \ref{messagingAndCommunication}, developer need not worry about AsyncTask or publish results per se. \middleware uses event-based mechanism to communicate between services and subscribers.
D) Activities need to interact with Services running on different processes or apps using IPC, so in this case the developer instantiates the Messenger class inside the Service and defines a Handler that responds to different types of Message objects, also, this Messenger shares an IBinder with a ServiceConnection object, allowing the Activity to send commands to the Service using Message objects;  \middleware uses service-oriented architecture combined with Message Broker and Resource Locator via Channel Adapters to interact between Activities and Services.
E) the Activity interacts with a Remote Service using AIDL (Android Interface Definition Language) where a RemoteService.Stub object returns an instance of the RemoteService to the ServiceConnection so it can then register callbacks that will monitor the service, then a handler is used to send/receive message objects that implement the interface Parcelable which is used for marshalling purposes; and 

Additionally, \middleware implements clean concurrency, without any overengineered Android Framework Api, by using pure thread based model. \ref{sec_concurrency} 
F) an Activity needs to read data from built-in sensors (e.g., Accelerometer) so it connects to a Service that implements the SensorEventListener, then it gets an instance of SensorManager to register itself and starts listening to particular sensor events. It is worth noting that these scenarios are even more complex since they require additional effort that we have omitted for the sake of simplicity (such as registering Services and BroadcastReceivers on AndroidManifest, allowing permissions, access to native libraries and hardware, etc.). 

\begin{figure}
\centering
  \includegraphics[width=\columnwidth]{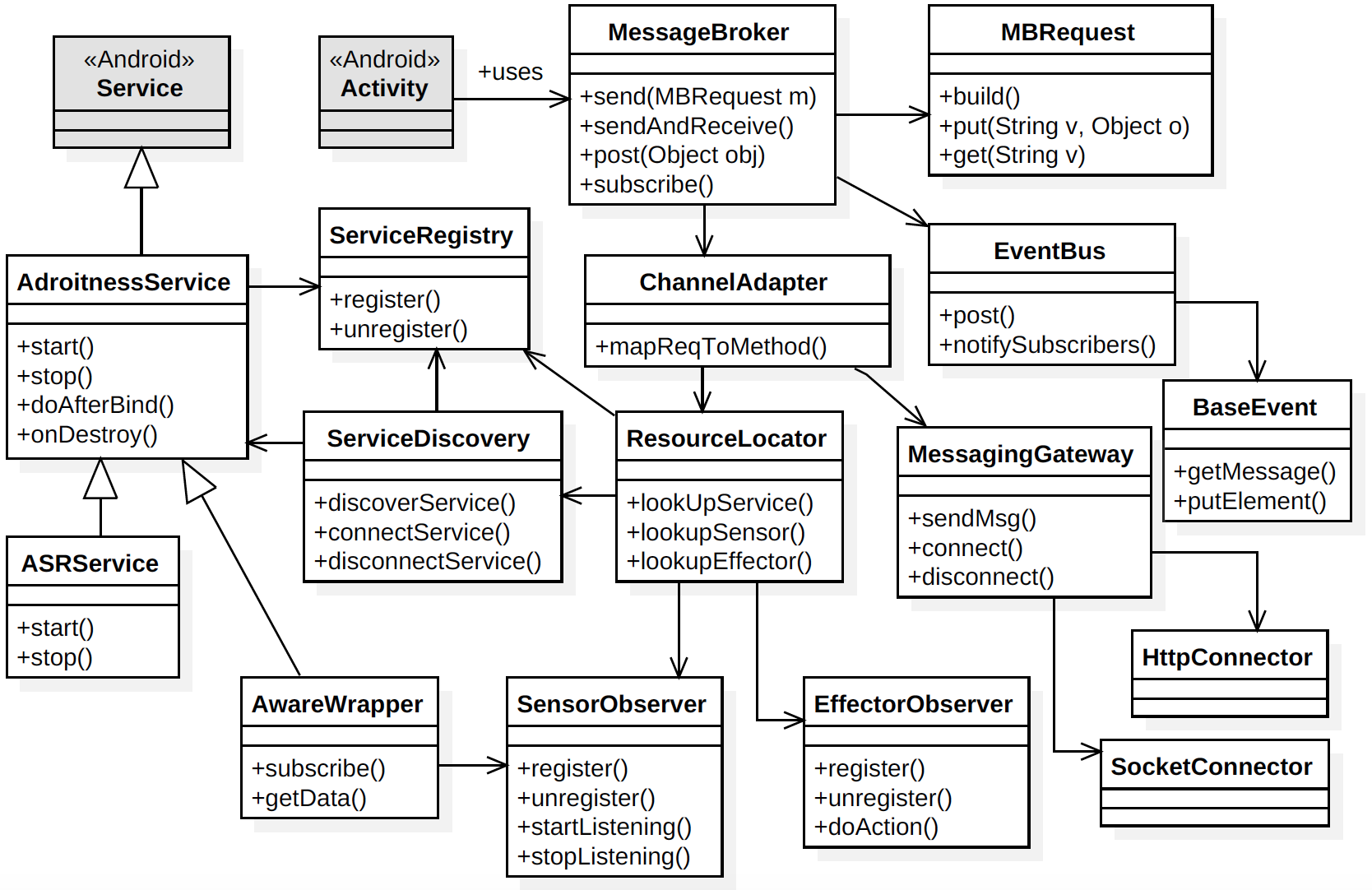}
  \caption{Adroitness Class Diagram.}~\label{fig:classes}
  \vspace{-1cm}
\end{figure}

\middleware implements sensors that make external changes and uses effectors that performs actions. \middleware abstracts away higher order functions to implement real-time scenarios such as detecting free-fall of phone. On the other hand, as we highlight the differences between AJF and \middleware , \middleware abstracts away the complexity of these 6 scenarios and simplify them to a single mechanism that connects Activities to the underlying Services, Sensors and Effectors (SSE) through a middleware layer that exposes only specific behavior to subscribe, post and receive messages to/from those components. Using the Clean Architecture principles for better separation of concerns and better modularization, \middleware allows to decouple the system into well-defined layers such as Presentation layer (i.e., Activities, GUI), Domain layer (i.e., business objects and rules) and Middleware layer (i.e., SSE). The Middleware is divided into four sub-layers: 1) a set of controllers that orchestrate the operation of SSE; 2) a Resource Manager that serves as a service discovery mechanism, resource locator, and dependency injector; 3) a Decision Rule Engine that creates synergies (rules) among those SSE; and 4) a communication layer composed by a Message Broker component and multiple Channel Adapters.
This layer uses minimal android dependencies, meaning that no Handlers, AsyncTasks, Messengers, BroadcastReceivers, Binders, nor ServiceConnections are used, instead, a lightweight but yet powerful concurrency and communication model is proposed, as described in further sections. It is worth noting that the only dependency between app GUI (Activity) and \middleware on the class diagram in Figure~\ref{fig:classes} is the MessageBroker, this centralization reduces complexity and increases maintainability. 


\subsection{Sensing and Acting} 
Sensors allow \middleware to detect external changes (e.g., variation in acceleration), user's events (e.g, gestures), and events among phones (e.g., phone1 notifies its proximity to phone2); whereas \emph{Effectors} perform actions as the result of making a decision (e.g, make a phone call). \middleware extends the Android SensorFramework (as described by scenario F in Figure~\ref{fig_architecture}.a) and adds high-order functions while abstracting away the atomic operations, e.g., the Accelerometer sensor is equipped with a mechanism for detecting free-fall so developers do not need to check whether the phone's 3-axis vector sum is equal to 0.
%


\subsection{Service-Orientation}  \label{sec_serviceOriented}
Since sensors' and effectors' extensibility is limited by phone's hardware, we enhanced them by using \middleware \emph{Services} (which extend Android Services), that is, application components that perform discrete functions either locally (in the phone) or remotely (on a server). Due to services extend far beyond the scope of phone's built-in sensors and effectors, they can be considered as cyber-sensors/effectors.
\middleware was designed based on a Service-Oriented Architecture (SOA) in order to promote loose coupling between services. We defined a Resource Manager pattern in charge of: 1)  maintaining a service registry which contains information about how to dispatch requests to services; 2) carrying out service discovery operations by using a resource locator pattern; 3) registering pluggable services that can be added or removed dynamically by using dependency injection; and 4) executing a Service Manager that controls the services lifecycle (start, destroy, bind, etc.). Our SOA architecture is empowered by the use of an event-driven mechanism that allows fast decoupled interaction between Android Services and Activities. 
\middleware provides a set of pre-defined pluggable services (e.g, weather, calendar, email, Automatic Speech Recognition -- ASR, access to Knowledge Bases, just to name a few, but developers can extend this set of services and add customized services that can be hooked into the middleware.

%

\subsection{Messaging and communication} \label{messagingAndCommunication}
\subsubsection{Message Broker} 
\middleware's architecture uses a Message Broker, an enterprise integration pattern that can negotiate and facilitate communication between a highly encapsulated set of services and UI components. It is in charge of routing, transforming, aggregating and decomposing messages. Using this pattern, our middleware provides a high level of abstraction, making transparent the communication between activities and services, that is, developers do not have to know the low level implementation details to interact with services, they only have to create a message broker request (MBRequest) instance and pass it to the message broker, then it will deliver the request to the corresponding service. This approach allows a clean way to maintain the code since all the interaction between components is centralized in the message broker. Finally, the message broker uses an event bus to communicate with services and activities through a publish/subscribe mechanism.

\subsubsection{Channel Adapter}
A Channel Adapter acts as a messaging client to the messaging system and invokes \middleware functions via a service-supplied interface. This way, any service can connect to the messaging system and be integrated with other services as long as it has a proper Channel Adapter.

\subsubsection{Messaging Gateway}

\begin{figure}
\centering
  \includegraphics[width=\columnwidth]{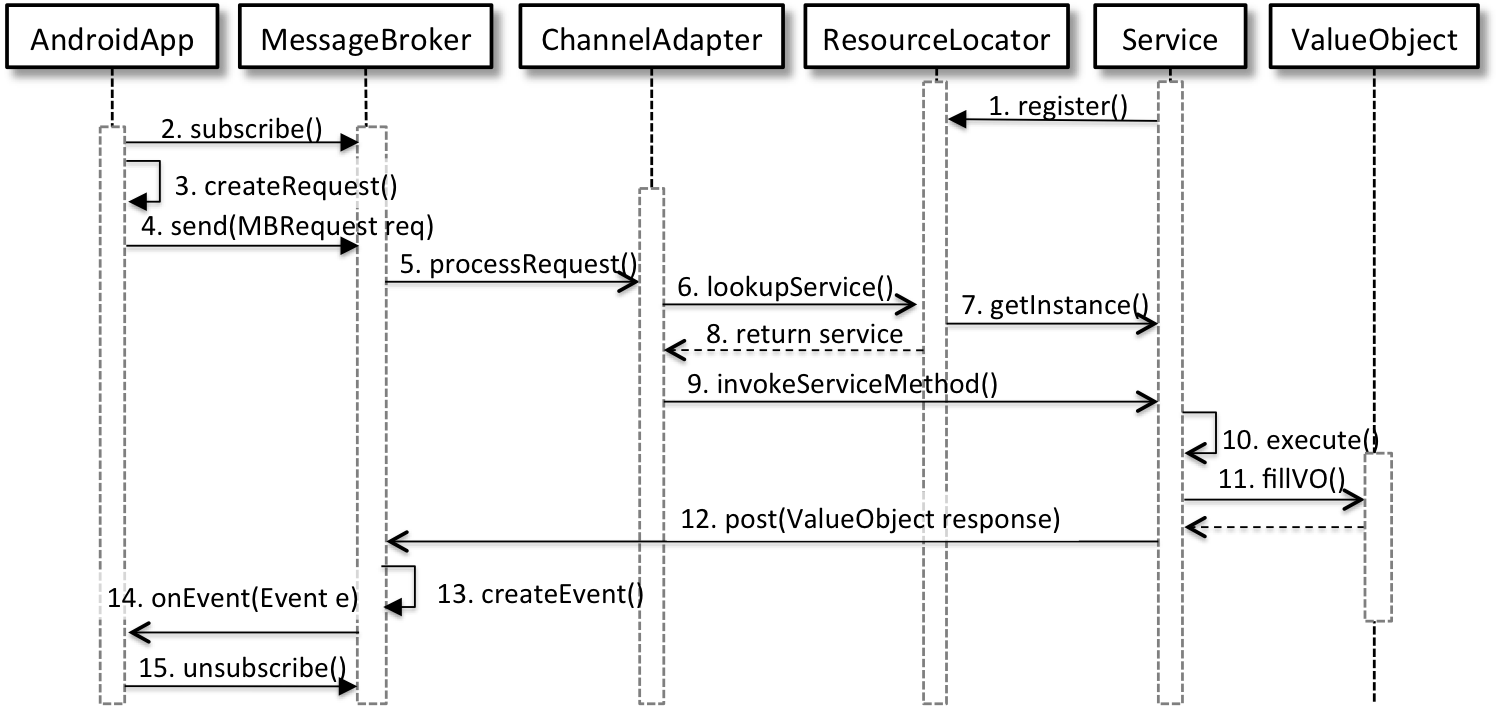}
  \vspace{-.5cm}
  \caption{Sequence Diagram for end-to-end message passing.}
  \vspace{-1cm}
  ~\label{fig:sequence}
\end{figure}

We used a Messaging Gateway to encapsulate messaging-specific code and separate it from the rest of the \middleware code. This way, only the Messaging Gateway code knows about the messaging system; the rest of the middleware code does not. This pattern allows \middleware to communicate with external applications using a broad variety of protocols (i.e, http, tcp sockets, rtsp, etc.) and data formats (e.g., json, xml, protobuffs, etc.). Figure~\ref{fig:sequence} depicts a simplified sequence diagram where the flow of messages within our middleware's components can be seen.

\subsection{Concurrency} \label{sec_concurrency}
In order to improve \middleware's latency footprint, throughput and interactivity, we defined a clean concurrency model that, on the one hand, radically eliminate the use of overengineered Android Framework constructs such as Handlers, AsyncTasks, ServiceConnections, Messengers, etc. and replace them with a pure Thread-based model that uses thread pools and async executors, being consistent with the clean architecture's principle that say: ``Architecture is About Intent, not Frameworks''~\cite{Martin:2017}; and on the other hand, uses a message-passing mechanism (the event bus) in order to pass messages between threads instead of sharing or accessing objects simultaneously, that way it is not necessary to protect the code by using locks, monitors and synchronized blocks that are computationally expensive. 

\subsection{Pluggability and Extensibility} 
\label{sec_cog_serv}
\begin{figure}
\centering
  \includegraphics[width=\columnwidth]{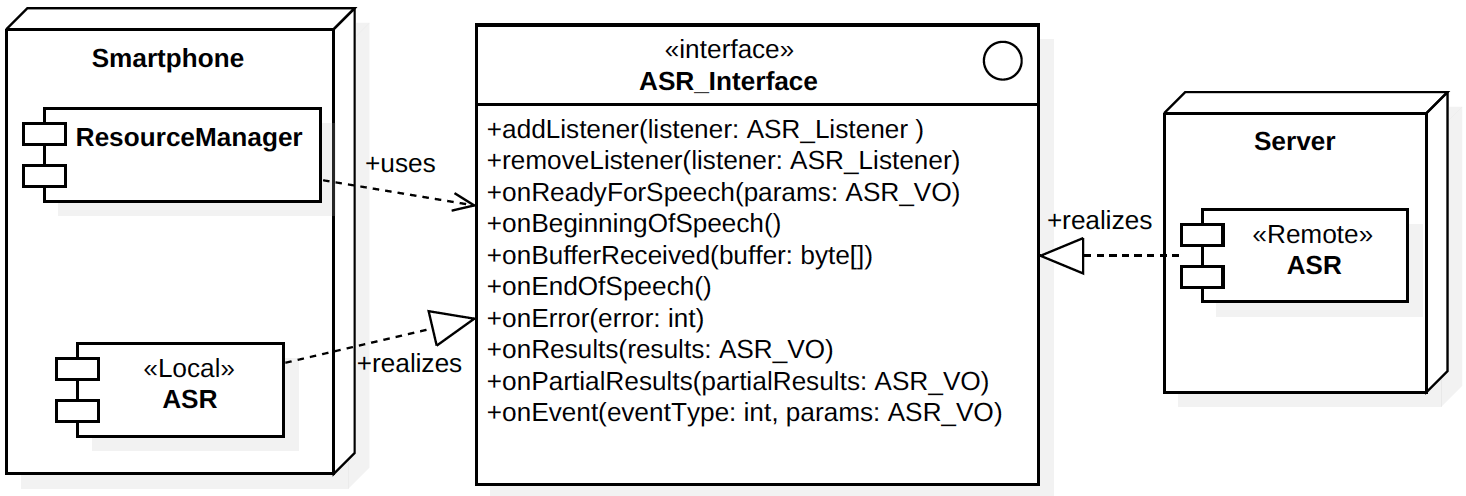}
  \vspace{-0.5cm}
  \caption{Simplified Component Diagram for ASR Service.}~\label{fig_components}
  \vspace{-0.8cm}
\end{figure}

Services are plug-in components based on a set of reusable contracts (interfaces) that are agreed to by developers so that these services can interoperate and reduce the development burden of common tasks. Each service component implements a common interface, with a uniform way to reference entities inside different components and across different naming schemes. For instance, take a look at the structural view on Figure~\ref{fig_components}. 
As you can see, there exist two implementations for the ASR (Automatic Speech Recognition) service component, one that runs on the phone (local) and one that runs on the server (remote). These two components realizes the common interface ASR\_Interface, which defines all the methods that both components have to implement. None component connects directly to another component, they do through interfaces. 
The ResourceManager dynamically bind a call against an interface to a specific implementation of the service at runtime based on the receiving parameters. This interface-centric design allows \middleware to be scalable and extensible. By letting each service component have multiple interfaces, we reduce the dependency of any one component on irrelevant features of another service component that it connects with. Also, adding new services incrementally can be accommodated more easily: developers can introduce new service components and add the relevant interfaces to existing ones. 
Different services can simply plug together, via defined interfaces for their services, to build higher-level behaviors (as we will explain in section~\ref{sec_rules}). This makes it easier to replace or upgrade parts; if CS components support the same (or compatible) interface, one part can be replaced by another.

\subsection{Decision Rule Engine} 
\label{sec_rules}
The Decision Rule Engine (DRE) is a rule-based system in charge of creating, validating, executing and specializing rules created by the user or developers. The DRE extends the \middleware's scope by aggregating different SSE, that is, services that can accomplish more complex processes and provide higher-level abstractions that allow developers to easily assemble entire use cases and interactions.
DRE is composed of a rule base (knowledge base); an inference engine in charge of the decision-making/reasoning loop that consists of three steps: matching the rule's left-hand sides against the contents of a working memory, determining the rule that must be chosen for execution through a conflict-resolution mechanism, and triggering the actions of the rule selected by the conflict-resolution mechanism; an a working memory that temporarily stores the content of partial solutions and the information provided by sensors and services.
A Rule is formally defined as a set of conditions so that \texttt{<left-side><operator><right-side>} and a set of actions so that \texttt{<event-action>:<component>: <method>:<params>}.
The DRE extends the \middleware's services scope by aggregating different services, sensors, and effectors, producing \emph{Composite Services} as a result, that is, services that can accomplish more complex processes and provide higher-level abstractions that allow developers to easily assemble entire use cases and interactions.
\begin{rule-listing}[Rules for a conversational interaction][single]
RULE: Rule1
  IF      Event.what equals Sensor.MIC.recording
  THEN    Event.post : Service.ASR.process : [MIC.bytes] AND
          Event.post : Service.NVB : processAF : [MIC.bytes]
RULE: Rule2
	IF      Event.what equals Service.ASR.response
	THEN    Event.post : Service.NLU : getIntent : [ASR.utterance]
RULE: Rule3
	IF      Event.what equals Service.NLU.intent
	THEN    Event.post : Service.DM : getIntent : [NLU.intent]
RULE: Rule4
	IF      Event.what equals Service.DM.intent
	THEN    Event.post : Service.NLG : realize : [DM.intent]
RULE: Rule5
	IF      Event.what equals Service.NLG.utterance
	THEN    Event.post : Service.TTS : realize : [NLG.utterance]
\end{rule-listing}

For instance, the five rules on Listing 1 represent a scenario for voice-based (conversational) mobile app, where: Rule1 is activated when the microphone sensor is recording user's voice and then triggers 2 services, the ASR and the Non-Verbal Behavior (NVB) that analyzes user's acoustic features (e.g., pitch, shimmer, jitter, etc.); Rule2 invokes the NLU (Natural Language Understanding) service when it receives the ASR response, that is, user's utterance transformed into text; Rule3 triggers the Dialog Manager (DM) service when the NLU has provided the intent; Rule4 invokes the NLG (Natural Language Generator) service once the DM has generated the system intent; and Rule5 synthesizes the utterance generated by NLG by triggering the TTS (Text-To-Speech) service. All this interaction is supported by an event-based mechanism. Rules do not make reference to specific component implementations (sensors, effectors, or services) but to generic contracts, that is, any component can be easily replaced by another component that accomplishes with the same contract. However, developers can add behaviors to reflect constraints, e.g., on Listing 2, when the microphone sensor starts recording, the DRE validates whether the WiFi network is off, if so, it will use an ASR implementation that runs locally and does not need internet connection (e.g., PocketSphinx), otherwise it will use a remote ASR (e.g., Google ASR).
\begin{rule-listing}[Rules for ASR activation]
RULE: Rule1
IF  Event.what equals Sensor.MIC.recording THEN
  IF   WiFi.turnedOn equals false
  THEN Event.post : Service.ASR : process : [ASR.Local,MIC.bytes]
  ELSE Event.post : Service.ASR : process : [ASR.Remote,MIC.byte]
\end{rule-listing}
\vspace{-0.2cm}
On Listing 3, we present a case where Rule1 is activated when DRE receives an event from the Facial Recognition (FR) service (given that camera sensor was previously activated), then it triggers the Emotion Recognition (ER) service, that in turn activates Rule2 which validates whether the inferred user's emotion is SAD; if so and if news service is running then the news feeds will be filtered to present only encouraging news (discarding disasters news, etc.), otherwise, if the Movie Recommendation (MR) service is open then it will rank the movies and recommend comedies first.
\begin{rule-listing}[Rules for creating composite behaviors based on FR and ER]
RULE: Rule1
	IF		Event.what equals Service.FR.response
	THEN 	Event.post : Service.ER.process : [Service.FR.AU]
RULE: Rule2
	IF 		Event.what equals Service.ER.response
	AND		Service.ER.Emotion equals Emotion.SAD 
	THEN    Event.post: UserModel : setEmotion : [SAD]
RULE: Rule3	
	IF      UserModel.emotion equals SAD
	AND     Service.NEWS.isOpen equals true
	THEN    Event.post : Service.NEWS : filter : [NEWS.Encouraging]
RULE: Rule4	
	IF      UserModel.emotion equals SAD
	AND     Service.MR.isOpen equals true
	THEN    Event.post : Service.MR : recommend : [MR.comedy]
\end{rule-listing}
Allowing users and developers to create rules that capture the behaviors they would like the phones to exhibit is not a new contribution, existing mobile applications such as IFTTT (IF This Then That~\cite{ovadia:2014}) allow end-users to create simple IF-THEN rules that link a set of triggers (e.g., sensors and web services) with specific actions (e.g., effectors and services). However, IFTTT only allows creating simple rules composed by one single condition and one single action, which results very limited when developers need to aggregate multiple sensors and services, which in turn may trigger multiple actions. Our DRE addresses this issue.

\vspace{-0.2cm}
\section{Implementation } \label{sec_implementation}
\vspace{-0.1cm}

\middleware's implementation is a mixture of an in-house development\footnote{See our GitHub repo: \textcolor{blue}{\url{https://github.com/ojrlopez27/adroitness-mobile}}} and the extension of third-party libraries such as GreenRobot's EventBus framework~\cite{eventbus:2017}, Google's Guava library, ZMQ and AWARE framework.

\subsection{Event-based Communication}
MessageBroker component extends GreenRobot's EventBus framework~\cite{eventbus:2017}, an Android optimized event bus that simplifies communication between Activities, Fragments, background Threads, Services, etc. by decoupling event senders and receivers, removing dependencies, and using a publisher/subscriber pattern for loose coupling. We extended GreenRobot Event Bus source code by adding new communication patterns such as Request/Response, Pair, and Router/Dealer. This allowed us to dispatch the events faster and avoid an additional logic necessary to filter the subscribers. We added an event-based mechanism also to intercept all message-passing so the Decision Rule Engine can make decisions based on message content.

\subsubsection{Caching} Android uses various techniques to share objects across services and activities such as SharedPreferences, Intents and Serializable/Parcealable Bundles. The potential disadvantages of these mechanisms are that they are computationally expensive transformations for serialization/deserialization that lead to an increased CPU processing for real-time transformations. We used Google Guava LRU (Least Recently Used) cache memory which holds object references in the memory without serializing/deserializing them. LruCache is thread-safe, improves the latency and reduces the time to access shared objects. 

\subsubsection{ZMQ messaging}
We used ZMQ messaging framework for communication with external servers. ZMQ is a high-performance asynchronous messaging library aimed to be used in distributed and concurrent applications with minimal latency footprint. Using this library, \middleware guarantees extremely low-latency responses, even when external servers, thanks to it access sockets directly. Using ZMQ, we could abstract away low-level communication details, such as dealing with different socket types, connection handling, framing, and even routing, so developers who is implementing a mobile app doesn't have to take care about communication issues.

\subsubsection{Access to Sensors and Effectors}

In order to abstract away details about how to access device's sensors and effectors, we extended the pool of plug-ins provided by AWARE~\cite{Ferreira:2013}, a middleware dedicated to instrument, infer, log and share mobile context information by capturing hardware-, software-, and human-based data. For instance, AWARE provides an effector for processing TTS outputs, however, it lacks of a sensor for processing ASR inputs. \middleware not only includes an effector for TTS but also provides a extensible API that allows to plug different kind of ASR implementations (e.g., Google Cloud ASR, Microsoft Bing Recognizer, CMU Pocket-sphinx, etc.) as well as an efficient streaming effector that streams out audio/video bytes to a remote server using RSTP (Real-Time Streaming Protocol) for those cases where user's audio features (e.g., pitch, jitter, shimmer, etc.) are processed in the server-side. Another example of how we extended AWARE is the LocationPlugin, which adds an abstraction layer on top of Google Fused Location, an energy efficient API that intelligently combines different signals (GPS and WiFi) to provide the location information needed. \middleware adds another abstraction layer on top of AWARE LocationPlugin (i.e., it exposes the same functionality to developers but using less code) and extends its functionality by keeping a location history (useful when no signal is available so the most recent location is used, or when some kind of inference based on regularities is required). Furthermore, \middleware uses Yahoo WOEID (Where On Earth ID) which provides additional information such as locality, timezone, county, town, state, etc.

\section{Evaluation} \label{sec_evaluation}

\begin{figure}
\includegraphics[width=\columnwidth] 
{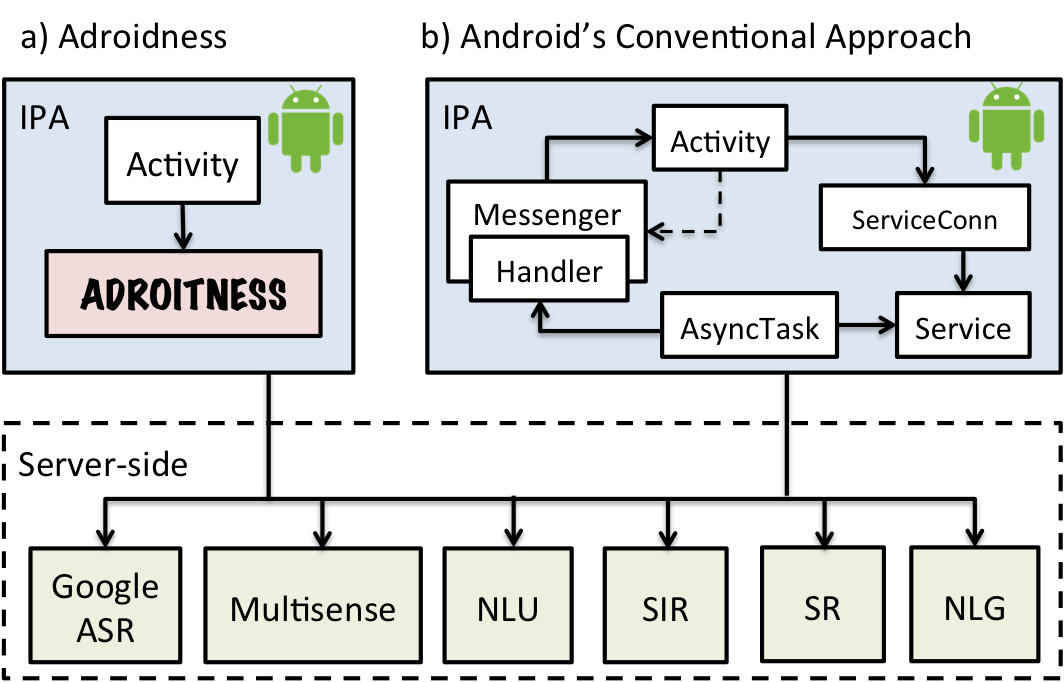}
\vspace{-0.5cm}
\caption{Test Scenario for Adroitness vs. AJF}
\vspace{-0.5cm}
\label{fig_scenario}
\end{figure}

In this section we present an empirical metric-based comparison between \middleware and the Android's Java Framework (AJF). We compared both approaches in the implementation of a conversational intelligent personal assistant that generated different kind of recommendations (e.g., movies, news, etc.) while keeping user's engagement through social dialogue. The \middleware approach was equipped with 7 different services: (Google ASR, Multisense for non-verbal behavior recognition~\cite{multisense:2017}, Microsoft NLU, NLG~\cite{socially:2016}
, a Social Intention Recognizer (SIR)~\cite{zhao:automatic}, a Social Reasoner (SR) for making decisions about conversational strategies~\cite{romero:2017}), and a Movie Recommendation System (MRS). 
For AJF we used 6 android's components: an ``Activity'' for the UI, a ``Service'' which connects to external servers, a ``ServiceConnection'' that monitors the service' state, an ``AsyncTask'' to perform background operations, and a ``Messenger+Handler'' to allow message-based communication across processes (see Figure~\ref{fig_scenario}).

\subsection{Measuring Latency and Performance} \label{sec_performanceanalysis}

In order to measure the latency of both \middleware and AJF, we conducted 15 experiments classified in 3 categories (Hardware configuration: XIAOMI Mi4C Smartphone, Android 5.1, Snapdragon 808, 64-bit Hexa Core 1.44GHz, and 16GB ROM.): using 1, 5 and 10 services. 
For each group, we collected data for sending/receiving 1, 10, 100, 1,000, and 10,000 messages. For latency experiments, we measured the time for sending and receiving 1, $10^1$, $10^2$, $10^3$ and $10^4$ messages. Each experiment was performed 10 times and then their harmonic mean was calculated, as shown in table \ref{latency_comparison}. Finally, a performance improvement rate between both approaches was estimated using the equation: $(AJF - \middleware)/AJF$. 
In general, \middleware's performance surpassed AJF's performance in a high rate when sending/receiving 1 message ($\cong95\%$) and then gradually decreased while the number of messages increased. It is worth noting that the performance rate was improved even more by our middleware when the number of services raised, and this pattern repeated across the experiments from 1 to 10K messages. For instance, when sending 10K messages using 1 single service, the difference between both approaches was almost insignificant (0.3\%), whereas when using 10 services to send 10K messages each one, \middleware improved the performance at a rate of up to $53\%$.
Finally, we ran a one-way ANOVA test to analyze the difference among the means of the 2 groups of data (\middleware vs. AJF), and given that the obtained p-value was less than the significance level ($p = 0.0032 \leq 0.5$) then we could conclude that there was a statistically significant difference between the 2 groups.
The performance experiments revealed a clear correlation between the number of services and the performance rate: the more services were running simultaneously the better \middleware performed. This is due to the multiple optimization levels of concurrency and communication among components: we drastically reduced the message-passing latency by using low-level thread manipulation and object caching instead of using Android-based components for communication (e.g., handlers, asynctasks, etc.) and for object-passing (e.g., serialization/deserialization of SharedPreferences and bundles)

\begin{table}[t]
  \caption{Latency Comparison. ADR = Adroitness, 1S = 1 service}
  \label{latency_comparison}
  \centering
  \resizebox{0.87\textwidth}{!}{
  \begin{minipage}{\textwidth}
  \begin{tabular}{ | l | r | r | r | r | r |}
    \hline
    {\small\textbf{Metric}}
    & {\small \textbf{1 Msg.}}
    & {\small \textbf{10 Msg.}}
    & {\small \textbf{100 Msg.}} 
    & {\small \textbf{1K Msg.}}
    & {\small \textbf{10K Msg.}}\\ \hline
    ADR 1S (ms)    & 1.3    &   14.4 &   104.4 &   671.2  &     5,337 \\ \hline
    AJF 1S (ms)     &  26    &   124  &   157   & 730 & 5,323 \\ \hline
    \rowcolor{Gray}
    Perf. Rate (\%) & 95 & 88.4 & 33.5     &   8.0   & -0.3 \\ \hline
    ADR 5S (ms) & 18.6 & 162.7 & 765.5 & 4,594 & 48,556 \\ \hline
    AJF 5S (ms) & 235 & 537 & 1,664.3 & 10,995 & 9,3343 \\ \hline
    \rowcolor{Gray}
    Perf. Rate (\%)        & 92.07 & 69.7 & 54.01 & 58.22 & 47.98 \\ \hline
    ADR 10S (ms) & 18.9 & 405 & 2,204 & 16,810 & 17,9289 \\ \hline
    AJF 10S (ms) & 631 & 1,523 & 5,196 & 37,696 & 38,7610 \\ \hline
    \rowcolor{Gray}
    Perf. Rate (\%) & 97 & 73.4 & 57.6 & 55.4 & 53.7 \\ \hline
  \end{tabular}
  \end{minipage}}
  \vspace{-0.5cm}
\end{table}
%

\subsubsection{Measuring Abstraction, Pluggability and Extensibility} \label{sec_metrics}

Quantitatively measuring software abstraction seems to be a non-trivial task. There exist different approaches in this regard, but the general consensus is that the more abstract an application is, the less complex and effort-consuming its development is~\cite{Glass:2002,Damasevicius:2006}. So our initial hypothesis was: \middleware should significantly decreased the complexity, size and effort to build the proposed scenario in comparison with AJF. 
We used the \emph{Cyclomatic Complexity} metric (CC), which is defined as the number of linearly independent paths within a graph that represents the source code flows, and is calculated as: $M = E − N + 2P$, where $E$ is the number of edges of the graph, $N$ is the number of nodes, and $P$ is the number of connected components~\cite{McCabe:1976}. Based on our analysis of measurements on Table~\ref{metric_comparison}, we deduced that both \middleware and AJF have low complexity (according to \cite{McCabe:1976} high complexity is over 15), however, the improvement rate demonstrated that \middleware reduces the complexity on $\cong30\%$ in comparison with AJF. 
We also measured the level of abstraction in terms of the minimum amount of implementation details that were exposed to the developer without loosing information content (the lesser exposed the most abstract). To this purpose, we used two MOOD metrics (Metrics for Object Oriented Design)~\cite{Abreu:1996}: the \emph{Method Hiding Factor} (MHF) and the \emph{Attribute Hiding Factor} (AHF) which were calculated across all classes in the system. It is worth noting that while our approach improved method hiding in a $78.67\%$, it only improved attribute hiding in a $4.8\%$, which is not particularly a significant difference, and this is due to we focused on creating high-level abstraction methods/classes while keeping the same attributes on both apps. 
In terms of size metric, we used Function Points (FP), a widely accepted industry standard (ISO/IEC 20926:2009) for functional sizing\footnote{Lines of Code is often criticized as ambiguous and meaningless~\cite{Damasevicius:2006}}. FP are units of measurement that express the amount of business functionality that an information system provides to a user. FP are estimated in terms of both data and transaction functions. As data functions, this metric estimates the amount and complexity of Internal Logical Files (ILF) and External Interface Files (EIF), and as transaction functions it estimates External Inputs (EI), External Outputs (EO) and External Inquiries (EQ). The corresponding FP's for each function have associated a complexity measure that can be Low (L), Average (A) or High (H). 
Based on the results presented on Table~\ref{metric_comparison}, we could observe that the main difference between both implementations was an increment of 5 FPs (for transaction functions) in AJF, which means that our approach reduced the amount of transactional functionality to be developed in $\cong36$\%. These difference of 5 FP's, when multiplied by the functional complexity factor (Low = 7), represents a 45\% of the total FP's for AJF, whereas it represents only a 23\% for \middleware.
The Total Function Point measure (TFP), which represents the total number of FPs after applying both an adjustment and a calibration factor, reflects that the whole app is $\cong53\%$ smaller in functionality \footnote{This means that less functionality has to be implemented to meet the same system's requirements, e.g., using reusable GUI components considerably reduces the amount of functionality to be implemented.} 
when using \middleware instead of AJF, which in turns represents a drop in effort in the same proportion.
\begin{figure}
\includegraphics[width=\columnwidth] 
{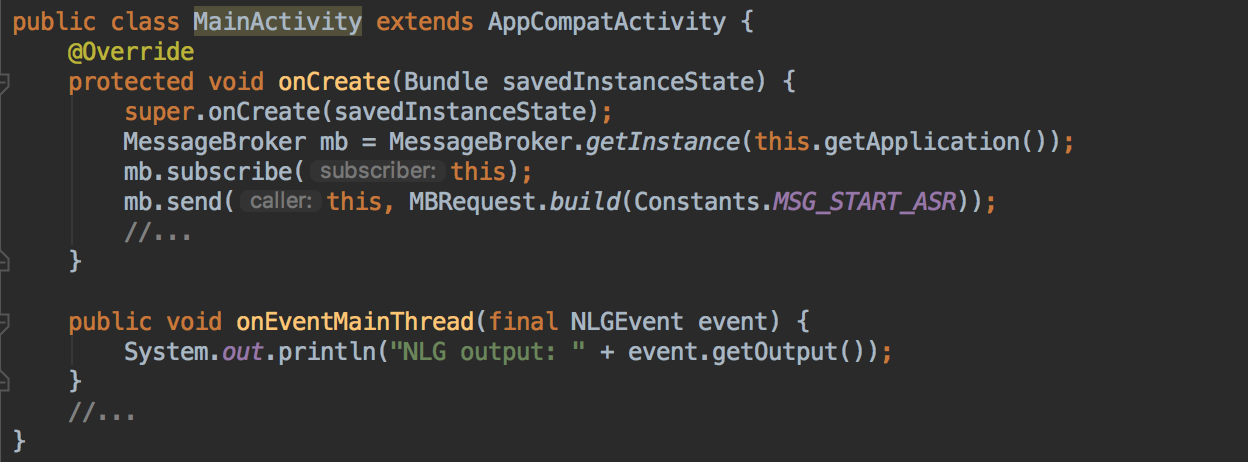}
\vspace{-0.5cm}
\caption{Code snippet for an end-to-end conversational interaction using \middleware. developers only had to write $\cong15$ lines of code (loc) for a two-step use case realization (i.e., in the 1st step, ASR is started on method onCreate and then, in the 2nd step, a NLG event that contains the utterance generated by the server is handled by onEventMainThread method), in comparison with more than 750 loc that the developers had to write in AJF for the same use case.}
\vspace{-0.7cm}
\label{fig_snippet}
\end{figure}
The Effort Person/Day (EPD) estimation is computed as $EPD = TFP/DR*DPM$, where DR is the delivery rate (in average, an android developer can implement 10 FPs per month~\cite{isbsg}) and DPM is days per person-month (21.5 business days per month). EPD can be better understood in terms of time and number of persons required to develop the app, let's say we have a team of 5 persons, using the AJF it would take 54.6 days (273.05/5) while using \middleware would take 25.8 days (129/5), which means a reduction of $\cong53\%$ of the required effort when using our approach (Figure~\ref{fig_snippet} shows a code snippet to illustrate how effort is decreased by hiding low-level implementation details). However, our empirical study revealed that development effort can be reduced more than 60\% when using our middleware, and this discrepancy with EPD may be due to unconsidered elements during the estimation.
\begin{table}
    \caption{Metrics Comparison of Adroitness (ADR) vs. AJF.}
    \label{metric_comparison}
    \centering
    \begin{tabular}{ | l | r | r | r | }
    \hline
    {\small\textbf{Metric}}
    & {\small \textbf{ADR}}
    & {\small \textbf{AJF}}
    & {\small \textbf{Improv. Rate}} \\ \hline
    CC	            & 1.2 		&	1.7		&	29.55\% \\ \hline
    MHF (\%)		& 42.04	& 	23.53	&	78.67\% \\ \hline
    AHF (\%)		& 97.79	&	93.33	&	4.78\% \\ \hline
    ILF + EIF (FP)	& 9 	&   9    	&   0\%	\\ \hline
    EI + EO + EQ (FP)	& 9		&   14  	&   35.71\%	 \\ \hline
    TFP (FP)		& 60	&   127 	&   52.75\%			\\ \hline
    EPD (person/day)   & 129 	&   273.05	    &	52.75\%		\\ \hline
    CBO		& 2.99		&	3.18	&	5.97\% \\ \hline
    CF (\%)	& 9 	    &   23.33	&	61.42\% \\ \hline
    LCOM	& 0.53		&	1.98	&	73.23\% \\ \hline
  \end{tabular}
  \vspace{-0.5cm}
\end{table}
Measuring pluggability and extensibility can be achieved by calculating the amount of coupling and cohesion in the system: the more loosely-coupled and high-cohesive the system is the more pluggable and extensible is. For this comparison we used 3 different metrics~\cite{Chidamber:1994}: Coupling between Object classes (CBO), Lack of Cohesion methods (LCOM), and Coupling factor (CF). In CBO, two classes are coupled when methods declared in one class use methods or instance variables of the other class; LCOM defines the number of different methods within a class that reference a given instance variable; and CF is the ratio of the maximum possible number of couplings in the system to the actual number of couplings not imputable to inheritance. The results of the measurements using these metrics are presented in Table~\ref{metric_comparison}. 
According to \cite{Chidamber:1994} a CBO $>$ 14 is too high, so the measurements suggest that both approaches have loose coupling, however, using CBO we could not conclude whether \middleware significantly improved the amount of coupling (it was only $5.97\%$), so we run the second metric for measuring coupling (CF) and we obtained a more significant difference between both approaches, this time, \middleware reduced the coupling between classes at a rate of $\cong62\%$. Also, CF should not exceed 12\%~\cite{Chidamber:1994} so AJF seems to be highly coupled (23.33\%) in comparison with our middleware (9\%). The main reasons why our approach lessen coupling are the use of dependency injection patterns and event-driven communication instead of direct invocations to classes. 
In LCOM metric, a result equals to 0 indicates a cohesive class, higher than 0 indicates that the class needs or can be split into two or more classes, since its variables belong in disjoint sets. Therefore, the results suggest that both approaches have certain lack of cohesion, however, \middleware seems to be $\cong73\%$ more cohesive than AJF. The main reason of this improvement is due to the architectural class decomposition into high-cohesive classes with a clear separation of concerns (e.g., communication, service discovery, reasoning, etc.). All the presented metrics are, in some way, biased and extremely sensible to the style of programming or errors during estimation, however, we fairly reduced the latter by using a reliable tool such as MetricsReloaded ~\cite{metrics:2010}.

\section{Related Work} 
\label{sec_related_work}
In this section, other work regarding the use of middleware on mobile operating systems is reviewed with respect to the requirements given in Section\ref{sec_motivation}. Besides, aforementioned AJF and \middleware specifications, architectural differences and evaluation, we categorize Middleware architectures into 3 categories:
\begin{enumerate}
    \item Cross-platform Middleware
    \item Component-Oriented Middleware
    \item Agent-Oriented Middleware
    \item Other types of Middleware
\end{enumerate}

\subsection{Cross-platform Middleware}
There are many 3rd party libraries and frameworks for Android indicating that the standard Android APIs are inadequate~\cite{Barnett:2015}. One alternative is to use cross-platform (hybrid) mobile frameworks based on  web technologies (JS, HTML, CSS). Cross-platform frameworks provide support to scripting languages such as JavaScript, TypeScript or Angular (e.g., Facebook ReactNative\cite{ReactNative:2018}, NativeScript\cite{NativeScript:2018} and Xamarin\cite{Xamarin:2018}) and some others use a web engine to render elements such as HTML5, CSS, and SVG, and execute the logic in a browser instance (e.g., JQuery Mobile\cite{JQueri:2018}, TheAppBuilder\cite{TheAppBuilder:2018}, and Apache PhoneGap\cite{ApacheCordova:2018}). Using cross-platform frameworks has advantages such as code sharing between the web and the app, leveraging developers current web language skills, and a plenty of open-source tools available. However, there are some disadvantages regarding fragmentation, compatibility, performance, UX issues, and memory, because they use a full web-rendering engine loaded just for the app and take a lot of GPU/CPU resources  increasing the app's response time~\cite{Jorgensen:2016}. 

\subsection{Component-Oriented Middleware}
Component-oriented middleware realizes the idea of interchangeable and reusable software components. They implement a component model, which defines syntax and semantics of component definitions and their relations\cite{Heineman:2001}. Several approaches, all based on OSGi\cite{OSGi}, have been proposed for the use on Android devices. Equinox was originally developed to provide a plugin-based architecture for the Eclipse IDE. In the progress of evaluating the application of Equinox on Android devices, necessary changes were added by\cite{Hargrave:2015}. For the time being, Equinox does not provide a concept for integrating UI. In consequence, it is uninteresting for many real-world scenarios. The Apache OSGi implementation Felix supports execution on Android since version 1.0.3. It is possible to use the Felix command line shell to add bundles and run console applications, just as with Equinox. Furthermore, Felix can be embedded in Android Apps and executed during the initialization of an app\cite{Felix:2015}. Based on this approach, Escoffier showed how to create Android apps that dynamically load .jar bundles\cite{Escoffier:2008}. The commercial OSGi implementation ProSyst mBS was designed for embedded hardware, and features explicit support for Android devices. As in Felix, application components are deployed as .jar bundles, which has to be implemented using the interface ApplicationFactory instead of Android activities. As opposed to the previously described OSGi implementations, the ProSyst platform is deployed inside a standalone Android application. To launch an individual application, a dummy app is installed on the device, instructing the platform application to load a specific application bundle\cite{Adjaz:2014}. This execution model enables sharing of the middleware platform between applications, while keeping the original user experience. Generally speaking, OSGi provides modularity and and pluggability features to Android-based apps, however, these implementations are tightly coupled to the AJF architecture, using Android components such as Activity, Service and Application factories, and Android-based communication protocols (e.g., AIDL -- Android Interface Definition Language), producing the same performance issues we described in our motivation section.

\subsection{Agent-Oriented Middleware}
Software agents provide a high-level approach to implement complex and concurrent software systems. In order to use such an abstraction, a runtime environment (platform) is required to provide services e.g., for executing or discovering agents. JaCa-Android\cite{Santi:2011} was specifically developed for Android and combines the Agents and Artifacts paradigm with an agent runtime called CArtAgO\cite{Alessandro:2007}. Agents are implemented using Jason, an AgentSpeak implementation\cite{Bordini:2007}. The runtime model is based on embedding the runtime platform into applications, including a central JaCa-Middleware application, which provides several artifacts to enable using services like contact management, localization or SMS from within agents. User interfaces are developed using default Android activities and are represented to the agents as artifacts to enable communication between them. The Agents and Artifacts approach allows for an elaborated integration of agent and Android design principles, but introduces an implementation language that is very different from traditional languages. Another agent-oriented middleware is JADE, which also features an Android version. Jade-Android can either integrate with a back-end or be executed as standalone platform. In any case, the runtime platform is included in applications; increasing the application sizes and loading times\cite{kalinowski:2015}. Agents can communicate with Android activities using the Object-to-Agent Interface (O2A). O2A utilizes Android intents sent by agents and received by activities as well as Java interfaces, which are used by activities to call agent methods\cite{Bergenti:2014}. The main issue with Agent-oriented approaches is that additional layers have to be embedded into the middleware (in the case of AgentSpeak and JADE, they support almost all the FIPA standard specifications -- The Foundation for Intelligent Physical Agents) resulting in a significant performance footprint.

\subsection{Android Architectural Components}
Android Architecture Components \cite{androidArchitectureComponents:2017} was introduced by Google Android in order to make make development efforts easier, address limited resources problems of Android devices and developing robust applications. For example life-cycle aware software or ViewModel or LiveData components \cite{androidArchitectureComponents:2017} basically add additional abstraction layers to user interface and data layers. The architecture components do not still address problems of poor concurrent performance, service composition, QoS. 

\section{Conclusions and Future Work} \label{sec_conclusions}
In this paper, we have presented an architectural middleware solution that supports the construction of robust, interactive, high-performance, production-quality mobile apps living on Android smartphone devices. Our main contributions can be summarized as follows: a) we used the clean architecture approach to guarantee a better separation of concerns and better modularization; b) we modeled specialized layers and modules with isolated responsibilities; c) the middleware architecture abstracts away the low-level design and implementation details such as communication, concurrency model, event handling, dependency injection, service discovery, among others; d) we improved the way Android Java Framework tackles quality attributes such as latency, extensibility, functional performance, and pluggability; e) latency was improved by avoiding the use of overengineered solutions to communicate Android components (e.g., Handlers, AsyncTasks, ServiceConnections, etc.) and replacing them by a lightweight threading model and a high-performance cache instead of using serialization/deserialization mechanisms; f) the architecture establishes reusable contracts for connecting (plug-in) service components that can be replaced any time; g) we defined a Decision Rule Engine that allows to create rules that binds sensors, effectors and services and facilitates the composition of more complex behaviors by aggregating multiple services; and h) we demonstrated that \middleware reduced the complexity, coupling, size, and effort of mobile apps implementation while improving the performance and cohesion.
Our future work will focus on several aspects: we are planning to make our system completely open-source, so the developer and research community can take advantage of the powerfulness of our middleware. The next steps in our development will be to use standards for flexibility, extensibility and pluggability such as OSGi (Open Service Gateway initiative). Also, we have identified the need of creating a semantic layer on top of our middleware in order to improve the service discovery process, provide more accurate and relevant information to higher-level layers, and make inferences about user's context. Also, we will implement a machine learning mechanism to discover and refine the rules orchestrated by DRE.

\section*{Acknowledgment}
The authors would like to thank the anonymous referees for their valuable comments and helpful suggestions.

\bibliographystyle{IEEEtranS}
\bibliography{main} 

\end{document}